\documentclass[submission,copyright,creativecommons]{eptcs}
\usepackage[utf8]{inputenc}
\usepackage{color}
\usepackage{amsmath, amssymb, amsthm}
\usepackage{subfig}
\usepackage{multirow}
\usepackage{tikz}
\usetikzlibrary{patterns}
\usetikzlibrary{arrows,shapes}
\tikzstyle{vertex}=[circle,minimum size=8pt,inner sep=0pt]
\tikzstyle{edge} = [draw,thick,->]
\tikzstyle{selected edge} = [draw,line width=5pt,line join=round,-,red!80]
\pgfdeclarelayer{background}
\pgfsetlayers{background,main}
\tikzstyle{sec vertex}=[circle,draw,minimum size=20pt,inner sep=0pt]
\tikzstyle{thd vertex}=[circle,draw,minimum size=8pt,inner sep=0pt]
\usepackage[ruled,vlined,linesnumbered,commentsnumbered,norelsize]{algorithm2e}

\providecommand{\event}{PDMC 2011} 

\title{Platform Dependent Verification:\\ On Engineering Verification Tools for 21st
  Century%
  \thanks{This research has been supported by the Czech Grant Agency grant
    No. GAP202/11/0312 and by Artemisia iFEST project grant No. 100203.}}
\author{
Lubo\v{s} Brim
\institute{Faculty of Informatics\\ Masaryk University\\ Brno, Czech Republic}
\email{brim@fi.muni.cz}
\and
Ji\v{r}\'{\i} Barnat
\institute{Faculty of Informatics\\ Masaryk University\\ Brno, Czech Republic}
\email{barnat@fi.muni.cz}
}
\def\titlerunning{Platform Dependent Verification: On Enineering Tools for 21st
  Century}
\def\authorrunning{L. Brim \& J. Barnat}

\newcommand{\BA}{\renewcommand{\baselinestretch}{1}\begin{algorithm}[!t]}
\newcommand{\EA}{\end{algorithm}\renewcommand{\baselinestretch}{1}}

\theoremstyle{remark}

\theoremstyle{definition}

\begin{document}
\maketitle

\section{Fighting the Space Explosion}

With the increase in complexity and degree of parallelism of computer systems,
it became even more important to develop formal methods for ensuring their
quality. Correctness and reliability became a must have flavor for business
success, and therefore, various techniques for automated and semi-automated
formal verification and analysis have been designed and successfully applied.
Formal verification and analysis bring many benefits such as early integration
in design process, more effective detection of logic errors, etc. Even though
introduction of formal analysis is rather costly, it pays off after all as it
results in significant reduction in verification time as well as development
costs and time-to-market. Attempts are being made to integrate formal
verification techniques and tools with other design approaches to support engineering of
complex industrial systems. The iFEST Artemisia project~\cite{ifest} is an
example of a promising tools integration in embedded systems domain.

\emph{Model checking} is a distinguished technique of formal verification of
complex hardware and software designs. Founders of the technique Edmund M. Clarke jr. (CMU, USA),
Allen E. Emerson (Texas at Austin, USA), and Joseph Sifakis (IMAG Grenoble,
France) were awarded ACM Turing Award in 2007 \emph{for their roles in
  developing model checking into a highly effective verification technology,
  widely adopted in the hardware and software industries}.  Unfortunately, model
checking procedure is computationally demanding and memory-intensive in general,
hence, its applicability to large and complex systems routinely seen in practice
these days is still limited. The major hampering factor is the \emph{state space
  explosion problem} due to which large industrial models cannot be efficiently
handled unless more sophisticated and scalable methods are used.

A lot of attention has been paid to the development of approaches to fight the
state space explosion problem~\cite{clarke01progress} in the field of automated
formal verification~\cite{pelanek09fighting}. Many techniques, such as  state
compaction~\cite{geldenhuys99runtime}, compression~\cite{spinbook}, state space
reduction~\cite{peled98ten,Symmetry1,Symmetry2}, symbolic state space
representation~\cite{burch92symbolic}, etc., were introduced to reduce the
memory requirements needed to handle the verification problem with a standard
sequential software tool.  These techniques allowed system developers to verify
larger systems without the need of increased computing power.

To verify even larger systems, however, no option was left out than to employ
combined computing power of multiple computing devices. Unfortunately, some
verification techniques cannot preserve their efficiency if adapted to
non-sequential models of computation, and therefore an urgent need for new and
quite different verification procedures emerged. Many  new techniques have
been introduced. Some of  them are applicable across a broad range of
computing platforms, some of them are tailored to the specific capabilities of a
particular hardware architectures. Examples include techniques to fight the
memory limits with an efficient utilization of external memory
devices~\cite{Stern98}, techniques that introduce cluster-based algorithms to
employ the aggregate power of network-interconnected
computers~\cite{stern97parallelizing,lerda99distributed,garavel01parallel,Bar04},
techniques to speed-up the verification process on multi-core
processors~\cite{holzmann07design,BBR08,Laarman10boosting}, etc.  

However, back
at the beginning of the 21st century, many of these techniques waited to be yet
discovered.
Even at that time the idea of using combined resources to increase the
computational power was far from being new in formal verification. Attempts to
use hard drives or parallel computers for verification of large systems have
appeared in the very early years of the automated formal verification
era. However, the inaccessibility of cheap parallel computers with sufficiently
fast external memory devices together with the negative theoretical complexity
results excluded these approaches from the main stream in formal
verification. Moreover, thanks to the Moore's law, the performance of software
tools kept improving continuously for years as the power of a single cored CPU
grew. The situation changed dramatically with oncoming of multi-core CPU
chips. The progress in computer design over the past decades had measured
several orders of magnitude with respect to various physical parameters such as
power consumption, efficiency, physical size or cost. As a result, it became
more efficient for chip producers to introduce multiple CPU cores on a single
chip rather than to increase the speed of a single core. As the speed of a
single core virtually stopped growing, every piece of software that was built
upon a serial algorithm could not take the advantage of technological progress
anymore. The focus of parallel and distributed-memory computing community
shifted away from unique massively parallel systems competing for world records
towards smaller and more cost effective systems built up from small and cheap
personal computer parts. Suddenly, the need for parallel processing become
rather general and wide spread in all science fields relying on complex
computation operations, automated formal verification being not an exception.
As a matter of fact, the interest in the \emph{platform-dependent} formal verification
has been revived.

\section{The DiVinE Story!}

DiVinE~\cite{BBC+06, BBCR10} is a tool for LTL model checking and reachability
analysis of discrete distributed systems. The tool is able to efficiently
exploit the aggregate computing power of multiple network-interconnected
multi-cored workstations in order to deal with extremely large verification
tasks. As such it allows to analyse systems of which size is far beyond the size of
systems that can be handled with regular sequential tools. DiVinE tool follows
the explicit-state automata-based approach to LTL model checking.  Due to Vardi
and Wolper~\cite{vardi86autom}, the LTL model checking problem reduces to the
problem of emptiness of B\"uchi automata, hence to the problem of the accepting
cycle detection in the underlying directed graph of a B\"uchi automaton.

\subsection{Parallel Algorithms in LTL Model Checking}

The need of parallel processing in automated formal verification stemmed from
the desire to fight the state space explosion problem by employing aggregate
memory of multiple network interconnected workstations. The crucial aspect
studied at first was how to distributed the work among participating processors
in order to take advantage of aggregate memory and parallel processing at the
same time.

Based on a parallel algorithm for state space
generation~\cite{caselli95parallel} a static partitioning scheme relying on a
hash function was introduced~\cite{ciardo98distributed}. As observed by multiple
researchers, the hash-based partitioning yields better space locality if only
parts of the state descriptor are used as the input to the partitioning
function. There were approaches requiring the user of the tool to specify the
concrete parts of the state descriptor to be used for
partitioning~\cite{ciardo98distributed,lerda99distributed}, other approaches
employed automated or semi-automated techniques to do
it~\cite{ciardo97automated}. DiVinE implicitly uses a hash-based partitioning
over the full state descriptor. Parts of the descriptor used for partitioning
might be statically redefined prior compilation. Techniques for load balancing the
set of visited states, also known as re-partitioning techniques, have been
suggested~\cite{allmaier97parallel,lerda01addressing,KumarM05} as well as state
space generation schemes employing probability
aspects~\cite{knottenbelt98probability,knottenbelt00probabilistic}. Nevertheless,
none of them have been implemented in DiVinE.

When DiVinE model checker development started, several previous tools had
existed. The first known public implementation of a distributed memory tool for
verification of communication protocols was the parallel implementation of the
Mur$\varphi$ tool~\cite{dill96themurphi,stern97parallelizing}. Mur$\varphi$
parallel work-flow relied on the standard MPI-like approach to messaging,
nevertheless, active messages were later introduced into Mur$\varphi$ to improve
its efficiency~\cite{eicken92active}. The successful story of the Mur$\varphi$
was followed by other verification tools:
SPIN~\cite{lerda99distributed,lerda01addressing}, CADP~\cite{garavel01parallel},
UPPALL~\cite{behrmann00distributed}, etc. Distributed-memory state space
generation as a technique of automated formal verification also appeared in the
context of Petri Nets~\cite{ciardo98distributed,Heljanko02parallelisation} and
Markov chains~\cite{haverkort98efficiency,haverkort99ontheefficient}.

Prior the DiVinE model checker, the existing distributed-memory parallel tools
were focused on state-space generation and reachability analyses only. The
reason was simple: the lack of parallel algorithms for accepting cycle detection
in distributed-memory setting. Nested Depth First Search algorithm (Nested DFS)
and other algorithms relying on dept-first search stack cannot be used in
distributed-memory setting as the distributed and parallel maintenance of the
depth-first search stack is inefficient~\cite{Bar04}. Therefore, new parallel
and distributed-memory algorithms for accepting cycle detection had to be
introduced with the development of DiVinE tool. The first implementation of
DiVinE employed the so called dependency structure~\cite{BBS01} to record the
reachability relation among accepting states of a distributed graph and applied
the topological sort algorithm~\cite{kahn62} to detect the presence of a
self-reachable accepting state. Other parallel algorithms appeared with the time
building upon various ideas: detection of negative cycles (NEGC
Algorithm)~\cite{BCKP-FSTTCS01,brim03parallel}, explicit-state implementation of
symbolic SCC hull detection (OWCTY Algorithm)~\cite{cerna03distributed}, value
propagation (MAP Algorithm)~\cite{brim04accepting}, or algorithm based on
back-level edges as produced by a breadth-first search procedure (BLEDGE
Algorithm)~\cite{BBC03,BBC05a}.  These new algorithms differed in theoretic
complexity as well as practical efficiency. After large experimental evaluation,
some of the algorithms were discontinued in DiVinE. The latest version of DiVinE
employs a combination of MAP and OWCTY algorithms by
defualt~\cite{BBR09a}. Table in Figure~\ref{fig:algorithms} gives complexities
and on-the-fly abilities of newly introduced parallel LTL model checking
algorithms.

\begin{figure}[t!]
  \centering
 {\renewcommand{\arraystretch}{1.3}
      \begin{tabular}{|l||c|c|c|}
        \hline
        & Complexity & Optimality & On-The-Fly \\
        \hline
        \hline
        \textbf{Nested DFS} & {\cal O}(N+M) & Yes & Yes \\
        \hline
        \textbf{OWCTY Algorithm} & & & \\
        \multicolumn{1}{|r||}{\quad general LTL properties} & {\cal O}(N.(N+M)) & No & No \\
        \multicolumn{1}{|r||}{\quad weak LTL properties} & {\cal O}(N+M) & Yes & No \\
        \hline
        \textbf{MAP Algorithm} &  {\cal O}(N.N.(N+M)) & No & Yes \\
        \hline
        \textbf{MAP-OWCTY Algorithm} & & & \\
        \multicolumn{1}{|r||}{\quad general LTL properties} & {\cal O}(N.(N+M)) & No & Yes \\
        \multicolumn{1}{|r||}{\quad weak LTL properties} & {\cal O}(N+M) & Yes & Yes \\
        \hline
        \textbf{BLEDGE Algorithm} &  {\cal O}(N.N.(N+M)) & No & Yes \\
        \hline
        \textbf{NEGC Algorithm} &  {\cal O}(N.N.(N+M)) & No & Yes \\
        \hline\end{tabular}
    }

\caption{Overview of complexities and on-the-fly processing ability of Nested DFS and
  parallel algorithms for accepting cycle detection.}
\label{fig:algorithms}
\end{figure}

Distributed-memory processing cannot fight the state space explosion problem
alone and must be combined with other techniques. One of the most successful
technique to fight the state space explosion in explicit-state model checking is
Partial Order Reduction~\cite{peled98ten}.  DiVinE is able to perform this
reduction, however, new topological sort proviso had to be developed in order to
maintain efficiency of parallel and distributed-memory
processing~\cite{BBR10b}. Another important algorithmic improvement relates
to classification of LTL formulas~\cite{cerna02relating}. For some classes of
LTL formulas (weak LTL) the parallel algorithms may by significantly
improved~\cite{BBC02}. With this observation the OWCTY algorithm can be improved
so that its complexity even meets the complexity of the optimal sequential Nested
DFS algorithm, see Table in Figure~\ref{fig:algorithms}.  However, this
algorithm suffers from not being an on-the-fly algorithm. Since the on-the-fly
verification is an important practical aspect, we have devised a modification of
this algorithm that allows for on-the-fly verification in most verification
instances~\cite{BBR09a}.

While DiVinE focuses on ``complete'' verification, parallel distributed-memory
``incomplete'' verification due to lossy state compaction has been introduced by
PReach tool~\cite{BBPE+10}.

\subsection{Algorithm Engineering for Parallel LTL Model Checking}

There is no doubt that without an appropriate parallel algorithm the LTL
model checking procedure cannot be successfully adapted to contemporary parallel
computing platforms. Nevertheless, the algorithm is not the only ingredient required. Even
the best algorithms in theory may not outperform good-but-not-optimal algorithms
that are equipped with platform-aware heuristics. This observation is even more
applicable to parallel processing where the scalability and absolute runtime
reduction are typically more valued achievements than theoretical optimality. To
that end there is another ingredient behind the development of parallel and
distributed-memory tool DiVinE -- \emph{Algorithm Engineering}.

\begin{quotation}
  \noindent
  \emph{Efforts must be made to ensure that promising algorithms discovered by the
  theory community are implemented, tested and refined to the point where they
  can be usefully applied in practice.}\\ \hspace*{1mm } \hfill
  {\footnotesize[Aho et al. [1997], Emerging Opportunities for Theoretical
    Computer Science] }
\end{quotation}

\noindent
In other words, characteristics of individual computing platforms must be taken
into account in order to obtain efficient implementations on these platforms.
In order to take the advantage of the processing power various platforms provide,
algorithm and data structure implementations need to be platform-dependent and
platform-aware.

\subsubsection{Parallelism in Distributed-Memory}
\label{sec:dm}
Distributed-memory parallel platform was the first platform that the DiVinE tool
was adapted to. The intention was to aggregate computational power and
distributed system memory of multiple network interconnected workstations
(clusters) in order to facilitate the verification of large model checking
instances~\cite{BBS01,Bar04}. The general idea of employing the
distributed-memory platform for execution of a parallel graph algorithm was, and
still is, as follows. The set of vertices of the graph to be processed is
partitioned among participating computation nodes using a static partitioning
function. When a computation node processes a vertex it enumerates all its immediate
successors  and checks them for their ownership. If a newly
generated vertex is local according to the partitioning function, it is pushed
to the local queue where it waits for further processing. In the other case a
network message is created containing the vertex and sent to the queue of the
owning computation node. With this work-flow there is a message generated with
every edge connecting vertices from different partitions of the graph. This is
where the theory is done, however, when it comes to the implementation there are
still numerous design choices to be made. Some of them are detailed for
individual computing platforms in the following subsections.

Message aggregation and buffering are the standard techniques in parallel
computing to alleviate the burden of network communication overhead.  Therefore,
DiVinE tool maintains buffers of messages to be sent to individual
network-attached computing nodes. In the first implementation of DiVinE, a
buffer was flushed (messages were sent to network) upon one of the following
situations: 1) buffer was explicitly flushed by the executed graph algorithm, 2)
maximal number of messages to fit the buffer has been reached, 3) the local
computing node was (otherwise) idle, and 4) messages in the buffer were too old.
Deep experimental evaluation, however, showed that the fourth condition is
completely ineffective in terms of network flow, while its checking is quite
expensive. After dropping the fourth rule for flushing of buffers DiVinE
significantly gained in performance.

\begin{figure}[t!]
  \centering
  \begin{tabular}{cc}
    \raisebox{-.16\linewidth}{
      \includegraphics[width=.5\linewidth]{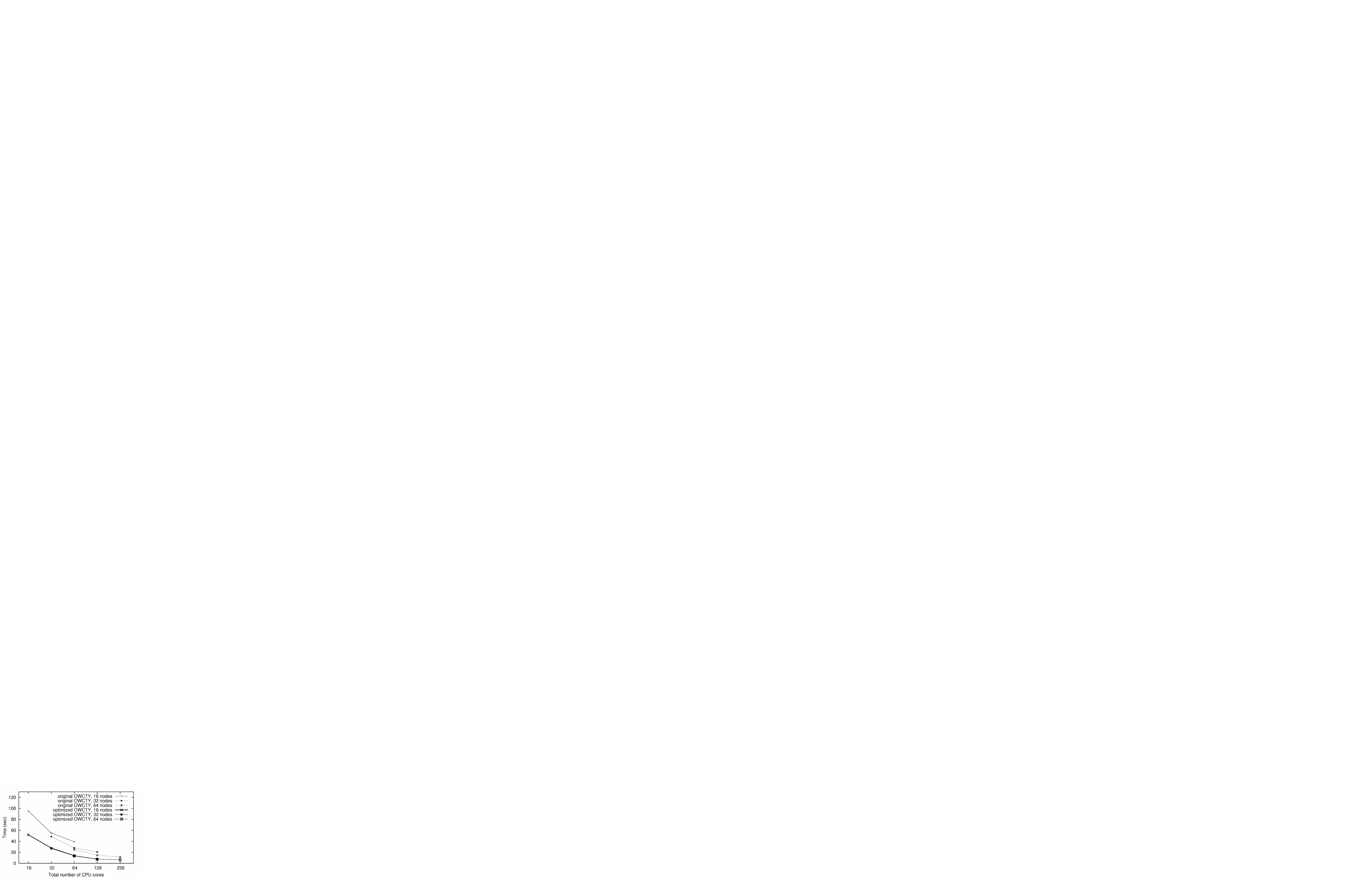}
    }
    & 
    \begin{tabular}{|r|r|r|}
      \hline
      Cores & Runtime (sec) & Efficiency \\
      \hline
      \hline
      1     &  631.7  & 100\% \\
      \hline
      64    &   13.3  &  74\% \\
      \hline
      128   &    7.4  &  67\% \\
      \hline
      256   &    5.0  &  49\% \\
      \hline
    \end{tabular}
    \end{tabular}
  \caption{DiVinE performance -- optimized ond original implementaion of OWCTY algorithm.}
  \label{fig:owcty}
\end{figure}

There were other distributed-memory performance bugs in earlier versions of
DiVinE. For exmaple, uncontrolled polling of incoming network messages, massive
flushing of all buffers at the same time, or insufficient separation of
initialization and computation phases. For more details
see~\cite{VBBB09}. Cumulative effect of elimination of these bugs from DiVinE is
shown in Figure~\ref{fig:owcty}.

\subsubsection{Parallelism in Shared-Memory}
\label{sec:sm}
Most techniques and results known from the distributed-memory setting are
straightforwardly applicable to shared-memory architectures.  DiVinE
architecture follows this observation, which means that if DiVinE is executed on a
multi-core machine with shared-memory, it mimics distributed-memory behavior. In
particular, the graph to be processed is partitioned among individual parallel
shared-memory threads in the same way as it would be in the distributed-memory
setting. Each individual thread  maintains its own hash table and its own
pool of vertices to be processed. Vertices belonging to different threads are
pushed to their local pools by means of lock-free shared-memory
queues~\cite{BBR07}. Relative advantages and disadvantages of shared versus
private hash tables  within the context of thread-private
pools of vertices to be processed have been discussed in~\cite{BR08}. These approaches were evaluated,
both theoretically and practically, in a prototype implementation~\cite{BBR08}.

Nevertheless, the scalability of parallel distributed-memory solutions to
shared-memory is often limited. Therefore, shared-memory specific techniques are
needed to improve the efficiency and scalability of existing parallel
distributed-memory solutions on shared-memory architectures. Examples of
successful shared-memory specific techniques include, e.g., shared communication
data structures~\cite{InggsB05,BBR07}, specific termination detection
techniques~\cite{BBR07}, dual-core algorithms~\cite{holzmann07design}, or quite
a unique partitioning scheme~\cite{Hol08}. As for  DiVinE, it
seems that the design choice of having  thread-private pools of vertices to be
processed  was not the best one~\cite{Laarman10boosting}. However, an experimental
confirmation  still waits to be done.

\subsubsection{Employing External Memory}
\label{sec:io}

Efficient algorithmic usage of computing devices with memory hierarchies is an
established research topic~\cite{ExternalMemory}. Numerous algorithms were
devised to efficiently utilize external-memory block devices, such as disks. The
efficiency of such an algorithm is typically measured in the number of I/O
(input/output) operations. To that end, the I/O efficient complexity has been
defined~\cite{Aggarwal88} and the standard breadth-first graph traversal
algorithms adapted to the I/O setting. The crucial technique used to do so is
the so called delayed duplicate detection~\cite{Vitter95} that has been further
improved in~\cite{ZhouH04a,BaoJ05,Weber06} and specialized for undirected
graphs~\cite{Korf04,KorfS05}. Regarding formal verification, the graph traversal
algorithms are used for state space generation and verification of safety
properties, see e.g. disk extension of the verification tool
Mur$\varphi$~\cite{Stern98,PennaITZ02}.

As for problems beyond the state space generation, breadth-first search graph
traversal algorithms are unsuitable. Therefore, the first  approach to
LTL model checking with external memory device employed a generic reduction of
the LTL model checking problem to the reachability
problem~\cite{BiereAS02}. Unfortunately, such a reduction results in a quadratic
grow in the memory demands, which effectively eliminates its application to large
scale industry cases. Therefore, ``incomplete'' verification approaches
 dominated the research field for some time. We have
seen random walks strategies implemented~\cite{JonesM04}, iterative deepening
and $A^*$ algorithms~\cite{JabbarE05,JabbarE06}, or breadth-first search based
approaches with limited amount of stored information~\cite{Lamborn08} to be
used.

The I/O branch of DiVinE was started with the invention of a new I/O algorithmic
technique that efficiently avoids the quadratic space overhead~\cite{BBS07}. The
new approach was further improved by introduction of the so called merge
omissions~\cite{BBSW08} that allowed for more efficient delayed duplicate
detection in the later stages of the computation. Various formulas for control
of what should be omitted were introduced~\cite{Evangelista08}, however, they
were not implemented within the I/O branch of DiVinE. A completely different
technique for trading time for space employing perfect hashing has been
implemented in the I/O branch of DiVinE. This technique is referred to as the
semi-external approach to LTL model checking problem~\cite{Semiext}.

\subsubsection{Many-Core Parallelism}
\label{sec:gpu}

After NVIDIA's CUDA technology~\cite{CUDA} was introduced, a lot of
computational demanding tasks have been accelerated by GPU-aware algorithms.
Examples of GPU accelerated procedures include, but are not limited to,
sorting~\cite{govindaraju06gputerasort}, reduce operations~\cite{cuda_reduce},
or numerous biological and physical simulations, such as protein
folding~\cite{Jayachandran06using}. As for the graph theory, successful
adaptation of general graph traversal algorithms have been reported
too~\cite{HarishN07,harish2009large} demonstrating the tremendous computational
power of the CUDA device. On the other hand, graphs to be explored efficiently
with a CUDA accelerated algorithm must be encoded explicitly in a compact way.

The CUDA technology as a computing platform attracted also researches in the
field of automated formal verification. The key challenge for which no
satisfactory solution is known yet is how to accelerate the generation of
explicitly encoded state space graph from the implicit definition. Preliminary
attempts to do so relate to explicit model checking. They suggest to
employ massively parallel check for enabled transitions emanating from the
vertices on the frontier of the search and their massive parallel
execution~\cite{ES08,SOCS09}.

Once the state space is generated and explicitly represented in an appropriate
sparse matrix like structure, many verification tasks can be accelerated using
CUDA technology. This has been successfully demonstrated, e.g., on verification
of probabilistic systems~\cite{BES09} or LTL model checking~\cite{BBCL09}.
Latest developments in DiVinE CUDA tool~\cite{BBC09} allow for efficient
utilization of multiple CUDA devices~\cite{BBBC10} and acceleration of detection
of strongly connected components~\cite{BBBC11}.

\section{Summary}

Platform dependent verification is an alternative approach how to make automated
formal verification attractive for industry. Despite significant progress in the
development of various specific techniques and tools on the algorithmic level,
mainly for parallel architectures, there is still a gap between pseudo-code and
implementation. Implementations must be tuned for specific platforms,
e.g. memory access patterns seem to play crucial role. In platform depended
verification we should learn to appreciate engineering solutions.

\bibliographystyle{eptcs}
\bibliography{pdmc11_invited}

\begin{thebibliography}{10}
\providecommand{\bibitemdeclare}[2]{}
\providecommand{\urlprefix}{Available at }
\providecommand{\url}[1]{\texttt{#1}}
\providecommand{\href}[2]{\texttt{#2}}
\providecommand{\urlalt}[2]{\href{#1}{#2}}
\providecommand{\doi}[1]{doi:\urlalt{http://dx.doi.org/#1}{#1}}
\providecommand{\bibinfo}[2]{#2}

\bibitemdeclare{article}{Aggarwal88}
\bibitem{Aggarwal88}
\bibinfo{author}{A.~Aggarwal} \& \bibinfo{author}{J.~S. Vitter}
  (\bibinfo{year}{1988}): \emph{\bibinfo{title}{The input/output complexity of
  sorting and related problems}}.
\newblock {\sl \bibinfo{journal}{Communications of the ACM}}
  \bibinfo{volume}{31}(\bibinfo{number}{9}), pp. \bibinfo{pages}{1116--1127}.

\bibitemdeclare{inproceedings}{allmaier97parallel}
\bibitem{allmaier97parallel}
\bibinfo{author}{S.~Allmaier}, \bibinfo{author}{S.~Dalibor} \&
  \bibinfo{author}{D.~Kreische} (\bibinfo{year}{1997}):
  \emph{\bibinfo{title}{{Parallel Graph Generation Algorithms for Shared and
  Distributed Memory Machines}}}.
\newblock In \bibinfo{editor}{G.~Bilardi}, \bibinfo{editor}{A.~G. Ferreira},
  \bibinfo{editor}{R.~L\"{u}ling} \& \bibinfo{editor}{J.~D.~P. Rolim}, editors:
  {\sl \bibinfo{booktitle}{Proceeding of the Parallel Computing Conference
  PARCO'97 (Bonn, Germany)}}, {\sl \bibinfo{series}{LNCS}}
  \bibinfo{volume}{1253}, \bibinfo{publisher}{Springer}, pp.
  \bibinfo{pages}{207--218}.
\newblock
  \urlprefix\url{http://citeseerx.ist.psu.edu/viewdoc/download?doi=10.1.1.17.2%
006&rep=rep1&type=pdf}.

\bibitemdeclare{inproceedings}{BaoJ05}
\bibitem{BaoJ05}
\bibinfo{author}{Tonglaga Bao} \& \bibinfo{author}{Michael Jones}
  (\bibinfo{year}{2005}): \emph{\bibinfo{title}{{Time-Efficient Model Checking
  with Magnetic Disk}}}.
\newblock In: {\sl \bibinfo{booktitle}{Tools and Algorithms for the
  Construction and Analysis of Systems (TACAS 2005)}}, {\sl
  \bibinfo{series}{LNCS}} \bibinfo{volume}{3440},
  \bibinfo{publisher}{Springer}, pp. \bibinfo{pages}{526--540},
  \doi{10.1007/978-3-540-31980-1\_34}.

\bibitemdeclare{phdthesis}{Bar04}
\bibitem{Bar04}
\bibinfo{author}{J.~Barnat} (\bibinfo{year}{2004}):
  \emph{\bibinfo{title}{{D}istributed {M}emory {LTL} {M}odel {C}hecking}}.
\newblock Ph.D. thesis, \bibinfo{school}{{M}asaryk {U}niversity {B}rno,
  {F}aculty of {I}nformatics}.
\newblock
  \urlprefix\url{http://anna.fi.muni.cz/papers/src/public/74925f5d8351a5a41120%
c11f26f3a21b.pdf}.

\bibitemdeclare{inproceedings}{BBBC10}
\bibitem{BBBC10}
\bibinfo{author}{J.~Barnat}, \bibinfo{author}{P.~Bauch},
  \bibinfo{author}{L.~Brim} \& \bibinfo{author}{M.~\v{C}e\v{s}ka}
  (\bibinfo{year}{2010}): \emph{\bibinfo{title}{{Employing Multiple CUDA
  Devices to Accelerate LTL Model Checking}}}.
\newblock In: {\sl \bibinfo{booktitle}{16th International Conference on
  Parallel and Distributed Systems (ICPADS 2010)}}, \bibinfo{publisher}{IEEE
  Computer Society}, pp. \bibinfo{pages}{259--266},
  \doi{10.1109/ICPADS.2010.82}.

\bibitemdeclare{inproceedings}{BBBC11}
\bibitem{BBBC11}
\bibinfo{author}{J.~Barnat}, \bibinfo{author}{P.~Bauch},
  \bibinfo{author}{L.~Brim} \& \bibinfo{author}{M.~\v{C}e\v{s}ka}
  (\bibinfo{year}{2011}): \emph{\bibinfo{title}{{Computing Strongly Connected
  Components in Parallel on CUDA}}}.
\newblock In: {\sl \bibinfo{booktitle}{International Parallel \& Distributed
  Processing Symposium (IPDPS'11)}}, \bibinfo{publisher}{IEEE Computer
  Society}, pp. \bibinfo{pages}{541--552}, \doi{10.1109/IPDPS.2011.59}.

\bibitemdeclare{inproceedings}{BBC02}
\bibitem{BBC02}
\bibinfo{author}{J.~Barnat}, \bibinfo{author}{L.~Brim} \&
  \bibinfo{author}{I.~{\v{C}}ern\'{a}} (\bibinfo{year}{2002}):
  \emph{\bibinfo{title}{Property driven distribution of {N}ested {DFS}}}.
\newblock In: {\sl \bibinfo{booktitle}{Proc. Workshop on Verification and
  Computational Logic}}, {\sl \bibinfo{series}{DSSE Technical Report}}
  \bibinfo{volume}{DSSE-TR-2002-5}, \bibinfo{organization}{University of
  Southampton, UK}, pp. \bibinfo{pages}{1--10}.
\newblock
  \urlprefix\url{http://anna.fi.muni.cz/papers/src/public/5f773c95a13c7b85f303%
123b36985210.pdf}.

\bibitemdeclare{inproceedings}{BBC03}
\bibitem{BBC03}
\bibinfo{author}{J.~Barnat}, \bibinfo{author}{L.~Brim} \&
  \bibinfo{author}{J.~Chaloupka} (\bibinfo{year}{2003}):
  \emph{\bibinfo{title}{{P}arallel {B}readth-{F}irst {S}earch {LTL}
  {M}odel-{C}hecking}}.
\newblock In: {\sl \bibinfo{booktitle}{18th IEEE International Conference on
  Automated Software Engineering (ASE'03)}}, \bibinfo{publisher}{IEEE Computer
  Society}, pp. \bibinfo{pages}{106--115}, \doi{10.1109/ASE.2003.1240299}.

\bibitemdeclare{article}{BBC05a}
\bibitem{BBC05a}
\bibinfo{author}{J.~Barnat}, \bibinfo{author}{L.~Brim} \&
  \bibinfo{author}{J.~Chaloupka} (\bibinfo{year}{2005}):
  \emph{\bibinfo{title}{{From Distributed Memory Cycle Detection to Parallel
  LTL Model Checking}}}.
\newblock {\sl \bibinfo{journal}{Electronic Notes in Theoretical Computer
  Science}} \bibinfo{volume}{133}(\bibinfo{number}{1}), pp.
  \bibinfo{pages}{21--39}, \doi{10.1016/j.entcs.2004.08.056}.

\bibitemdeclare{inproceedings}{BBR07}
\bibitem{BBR07}
\bibinfo{author}{J.~Barnat}, \bibinfo{author}{L.~Brim} \&
  \bibinfo{author}{P.~Ro\v{c}kai} (\bibinfo{year}{2007}):
  \emph{\bibinfo{title}{Scalable Multi-core LTL Model-Checking}}.
\newblock In: {\sl \bibinfo{booktitle}{Model Checking Software}}, {\sl
  \bibinfo{series}{LNCS}} \bibinfo{volume}{4595},
  \bibinfo{publisher}{Springer}, pp. \bibinfo{pages}{187--203},
  \doi{10.1007/978-3-540-73370-6\_13}.

\bibitemdeclare{inproceedings}{BBR08}
\bibitem{BBR08}
\bibinfo{author}{J.~Barnat}, \bibinfo{author}{L.~Brim} \&
  \bibinfo{author}{P.~Ro\v{c}kai} (\bibinfo{year}{2008}):
  \emph{\bibinfo{title}{{DiVinE Multi-Core -- A Parallel LTL Model-Checker}}}.
\newblock In: {\sl \bibinfo{booktitle}{Automated Technology for Verification
  and Analysis (ATVA 2008)}}, {\sl \bibinfo{series}{LNCS}}
  \bibinfo{volume}{5311}, \bibinfo{publisher}{Springer}, pp.
  \bibinfo{pages}{234--239}, \doi{10.1007/978-3-540-88387-6\_20}.

\bibitemdeclare{inproceedings}{BBR09a}
\bibitem{BBR09a}
\bibinfo{author}{J.~Barnat}, \bibinfo{author}{L.~Brim} \&
  \bibinfo{author}{P.~Ro\v{c}kai} (\bibinfo{year}{2009}):
  \emph{\bibinfo{title}{{A Time-Optimal On-the-Fly Parallel Algorithm for Model
  Checking of Weak LTL Properties}}}.
\newblock In: {\sl \bibinfo{booktitle}{Formal Methods and Software Engineering
  (ICFEM 2009)}}, {\sl \bibinfo{series}{LNCS}} \bibinfo{volume}{5885},
  \bibinfo{publisher}{Springer}, pp. \bibinfo{pages}{407--425},
  \doi{10.1007/978-3-642-10373-5\_21}.

\bibitemdeclare{inproceedings}{BBR10b}
\bibitem{BBR10b}
\bibinfo{author}{J.~Barnat}, \bibinfo{author}{L.~Brim} \&
  \bibinfo{author}{P.~Ro\v{c}kai} (\bibinfo{year}{2010}):
  \emph{\bibinfo{title}{{Parallel Partial Order Reduction with Topological Sort
  Proviso}}}.
\newblock In: {\sl \bibinfo{booktitle}{Software Engineering and Formal Methods
  (SEFM 2010)}}, \bibinfo{publisher}{IEEE Computer Society Press}, pp.
  \bibinfo{pages}{222--231}, \doi{10.1109/SEFM.2010.35}.

\bibitemdeclare{inproceedings}{BBS01}
\bibitem{BBS01}
\bibinfo{author}{J.~Barnat}, \bibinfo{author}{L.~Brim} \&
  \bibinfo{author}{J.~St\v{r}\'{\i}brn\'{a}} (\bibinfo{year}{2001}):
  \emph{\bibinfo{title}{Distributed {L}{T}{L} {M}odel-{C}hecking in
  {S}{P}{I}{N}}}.
\newblock In: {\sl \bibinfo{booktitle}{Proc. SPIN Workshop on Model Checking of
  Software}}, {\sl \bibinfo{series}{LNCS}} \bibinfo{volume}{2057},
  \bibinfo{publisher}{Springer}, pp. \bibinfo{pages}{200--216},
  \doi{10.1007/3-540-45139-0\_13}.

\bibitemdeclare{inproceedings}{BBC+06}
\bibitem{BBC+06}
\bibinfo{author}{J.~Barnat}, \bibinfo{author}{L.~Brim},
  \bibinfo{author}{I.~\v{C}ern\'{a}}, \bibinfo{author}{P.~Moravec},
  \bibinfo{author}{P.~Ro\v{c}kai} \& \bibinfo{author}{P.~\v{S}ime\v{c}ek}
  (\bibinfo{year}{2006}): \emph{\bibinfo{title}{{DiVinE -- A Tool for
  Distributed Verification (Tool Paper)}}}.
\newblock In: {\sl \bibinfo{booktitle}{{Computer Aided Verification}}}, {\sl
  \bibinfo{series}{LNCS}} \bibinfo{volume}{4144/2006},
  \bibinfo{publisher}{Springer Berlin / Heidelberg}, pp.
  \bibinfo{pages}{278--281}, \doi{10.1007/11817963\_26}.

\bibitemdeclare{article}{BBC09}
\bibitem{BBC09}
\bibinfo{author}{J.~Barnat}, \bibinfo{author}{L.~Brim} \&
  \bibinfo{author}{M.~\v{C}e\v{s}ka} (\bibinfo{year}{2009}):
  \emph{\bibinfo{title}{{DiVinE-CUDA: A Tool for GPU Accelerated LTL Model
  Checking}}}.
\newblock {\sl \bibinfo{journal}{Electronic Proceedings in Theoretical Computer
  Science (PDMC 2009)}} \bibinfo{volume}{14}, pp. \bibinfo{pages}{107--111},
  \doi{10.4204/EPTCS.14.8}.

\bibitemdeclare{inproceedings}{BBCL09}
\bibitem{BBCL09}
\bibinfo{author}{J.~Barnat}, \bibinfo{author}{L.~Brim},
  \bibinfo{author}{M.~\v{C}e\v{s}ka} \& \bibinfo{author}{T.~Lamr}
  (\bibinfo{year}{2009}): \emph{\bibinfo{title}{{CUDA accelerated LTL Model
  Checking}}}.
\newblock In: {\sl \bibinfo{booktitle}{15th International Conference on
  Parallel and Distributed Systems (ICPADS 2009)}}, \bibinfo{publisher}{IEEE
  Computer Society}, pp. \bibinfo{pages}{34--41}, \doi{10.1109/ICPADS.2009.50}.

\bibitemdeclare{inproceedings}{BBCR10}
\bibitem{BBCR10}
\bibinfo{author}{J.~Barnat}, \bibinfo{author}{L.~Brim},
  \bibinfo{author}{M.~\v{C}e\v{s}ka} \& \bibinfo{author}{P.~Ro\v{c}kai}
  (\bibinfo{year}{2010}): \emph{\bibinfo{title}{{DiVinE: Parallel Distributed
  Model Checker (Tool paper)}}}.
\newblock In: {\sl \bibinfo{booktitle}{Parallel and Distributed Methods in
  Verification and High Performance Computational Systems Biology (HiBi/PDMC
  2010)}}, \bibinfo{publisher}{IEEE}, pp. \bibinfo{pages}{4--7},
  \doi{10.1109/PDMC-HiBi.2010.9}.

\bibitemdeclare{inproceedings}{BBS07}
\bibitem{BBS07}
\bibinfo{author}{J.~Barnat}, \bibinfo{author}{L.~Brim} \&
  \bibinfo{author}{P.~\v{S}ime\v{c}ek} (\bibinfo{year}{2007}):
  \emph{\bibinfo{title}{{I/O Efficient Accepting Cycle Detection}}}.
\newblock In: {\sl \bibinfo{booktitle}{Computer Aided Verification}}, {\sl
  \bibinfo{series}{LNCS}} \bibinfo{volume}{4590},
  \bibinfo{publisher}{Springer}, pp. \bibinfo{pages}{281--293},
  \doi{10.1007/978-3-540-73368-3\_32}.

\bibitemdeclare{inproceedings}{BBSW08}
\bibitem{BBSW08}
\bibinfo{author}{J.~Barnat}, \bibinfo{author}{L.~Brim},
  \bibinfo{author}{P.~\v{S}ime\v{c}ek} \& \bibinfo{author}{M.~Weber}
  (\bibinfo{year}{2008}): \emph{\bibinfo{title}{{Revisiting Resistance Speeds
  Up I/O-Efficient LTL Model Checking}}}.
\newblock In: {\sl \bibinfo{booktitle}{Tools and Algorithms for the
  Construction and Analysis of Systems (TACAS).}}, {\sl \bibinfo{series}{LNCS}}
  \bibinfo{volume}{4963}, \bibinfo{publisher}{Springer}, pp.
  \bibinfo{pages}{48--62}, \doi{10.1007/978-3-540-78800-3\_5}.

\bibitemdeclare{article}{BR08}
\bibitem{BR08}
\bibinfo{author}{J.~Barnat} \& \bibinfo{author}{P.~Ro\v{c}kai}
  (\bibinfo{year}{2008}): \emph{\bibinfo{title}{{Shared Hash Tables in Parallel
  Model Checking}}}.
\newblock {\sl \bibinfo{journal}{ENTCS}}
  \bibinfo{volume}{198}(\bibinfo{number}{1}), pp. \bibinfo{pages}{79--91},
  \doi{10.1016/j.entcs.2007.10.021}.

\bibitemdeclare{inproceedings}{behrmann00distributed}
\bibitem{behrmann00distributed}
\bibinfo{author}{G.~Behrmann}, \bibinfo{author}{T.~S. Hune} \&
  \bibinfo{author}{F.~W. Vaandrager} (\bibinfo{year}{2000}):
  \emph{\bibinfo{title}{Distributed Timed Model Checking --- How the Search
  Order Matters}}.
\newblock In: {\sl \bibinfo{booktitle}{Proc. 12th Conference on Computer-Aided
  Verification CAV00}}, {\sl \bibinfo{series}{LNCS}} \bibinfo{volume}{1855},
  \bibinfo{publisher}{Springer}, pp. \bibinfo{pages}{216--231},
  \doi{10.1007/10722167\_19}.

\bibitemdeclare{article}{BiereAS02}
\bibitem{BiereAS02}
\bibinfo{author}{Armin Biere}, \bibinfo{author}{Cyrille Artho} \&
  \bibinfo{author}{Viktor Schuppan} (\bibinfo{year}{2002}):
  \emph{\bibinfo{title}{{Liveness Checking as Safety Checking}}}.
\newblock {\sl \bibinfo{journal}{Electronic Notes in Theoretical Computer
  Science}} \bibinfo{volume}{66}(\bibinfo{number}{2}), pp. \bibinfo{pages}{160
  -- 177}, \doi{10.1016/S1571-0661(04)80410-9}.

\bibitemdeclare{inproceedings}{BBPE+10}
\bibitem{BBPE+10}
\bibinfo{author}{Brad Bingham}, \bibinfo{author}{Jesse Bingham},
  \bibinfo{author}{Flavio~M. de~Paula}, \bibinfo{author}{John Erickson},
  \bibinfo{author}{Gaurav Singh} \& \bibinfo{author}{Mark Reitblatt}
  (\bibinfo{year}{2010}): \emph{\bibinfo{title}{{Industrial Strength
  Distributed Explicit State Model Checking}}}.
\newblock In: {\sl \bibinfo{booktitle}{Parallel and Distributed Methods in
  Verification, 2010 Ninth International Workshop on, and High Performance
  Computational Systems Biology, Second International Workshop on}},
  \bibinfo{publisher}{IEEE Computer Society}, pp. \bibinfo{pages}{28--36},
  \doi{10.1109/PDMC-HiBi.2010.13}.

\bibitemdeclare{inproceedings}{BES09}
\bibitem{BES09}
\bibinfo{author}{D.~Bosnacki}, \bibinfo{author}{S.~Edelkamp} \&
  \bibinfo{author}{D.~Sulewski} (\bibinfo{year}{2009}):
  \emph{\bibinfo{title}{{Efficient Probabilistic Model Checking on General
  Purpose Graphics Processors}}}.
\newblock In: {\sl \bibinfo{booktitle}{Model Checking Software (SPIN 2009)}},
  {\sl \bibinfo{series}{LNCS}} \bibinfo{volume}{5578},
  \bibinfo{publisher}{Springer}, pp. \bibinfo{pages}{32--49},
  \doi{10.1007/978-3-642-02652-2\_7}.

\bibitemdeclare{techreport}{brim03parallel}
\bibitem{brim03parallel}
\bibinfo{author}{L.~Brim}, \bibinfo{author}{I.~{\v{C}}ern\'a} \&
  \bibinfo{author}{L.~Hejtm\'anek} (\bibinfo{year}{2003}):
  \emph{\bibinfo{title}{Parallel {A}lgorithms for {D}etection of {N}egative
  {C}ycles}}.
\newblock \bibinfo{type}{Technical Report} \bibinfo{number}{FIMU-RS-2003-04},
  \bibinfo{institution}{Faculty of Informatics, Masaryk University Brno}.
\newblock
  \urlprefix\url{http://www.fi.muni.cz/informatics/reports/files/2003/FIMU-RS-%
2003-04.pdf}.

\bibitemdeclare{inproceedings}{brim04accepting}
\bibitem{brim04accepting}
\bibinfo{author}{L.~Brim}, \bibinfo{author}{I.~{\v{C}}ern\'a},
  \bibinfo{author}{P.~Moravec} \& \bibinfo{author}{J.~{\v{S}}im\v{s}a}
  (\bibinfo{year}{2004}): \emph{\bibinfo{title}{Accepting Predecessors are
  Better than Back Edges in Distributed {LTL} Model-Checking}}.
\newblock In: {\sl \bibinfo{booktitle}{Formal Methods in Computer Aided Design
  (FMCAD)}}, {\sl \bibinfo{series}{LNCS}} \bibinfo{volume}{4144},
  \bibinfo{publisher}{Springer}, pp. \bibinfo{pages}{352--366},
  \doi{10.1007/978-3-540-30494-4\_25}.

\bibitemdeclare{inproceedings}{BCKP-FSTTCS01}
\bibitem{BCKP-FSTTCS01}
\bibinfo{author}{L.~Brim}, \bibinfo{author}{I.~\v{C}ern\'{a}},
  \bibinfo{author}{P.~Kr\v{c}\'{a}l} \& \bibinfo{author}{R.~Pel\'{a}nek}
  (\bibinfo{year}{2001}): \emph{\bibinfo{title}{{Distributed LTL Model Checking
  Based on Negative Cycle Detection}}}.
\newblock In: {\sl \bibinfo{booktitle}{Proc. of Foundations of Software
  Technology and Theoretical Computer Science (FST TCS 2001)}}, {\sl
  \bibinfo{series}{LNCS}} \bibinfo{volume}{2245},
  \bibinfo{publisher}{Springer}, pp. \bibinfo{pages}{96--107},
  \doi{10.1007/3-540-45294-X\_9}.

\bibitemdeclare{article}{burch92symbolic}
\bibitem{burch92symbolic}
\bibinfo{author}{J.~R. Burch}, \bibinfo{author}{E.~M. Clarke},
  \bibinfo{author}{K.~L. McMillan}, \bibinfo{author}{D.~L. Dill} \&
  \bibinfo{author}{L.~J. Hwang} (\bibinfo{year}{1992}):
  \emph{\bibinfo{title}{Symbolic Model Checking: $10^{20}$ States and Beyond}}.
\newblock {\sl \bibinfo{journal}{Information and Computation}}
  \bibinfo{volume}{98}(\bibinfo{number}{2}), pp. \bibinfo{pages}{142--170}.
\newblock \urlprefix\url{http://citeseer.nj.nec.com/burch92symbolic.html}.

\bibitemdeclare{inproceedings}{caselli95parallel}
\bibitem{caselli95parallel}
\bibinfo{author}{S.~Caselli}, \bibinfo{author}{G.~Conte} \&
  \bibinfo{author}{P.~Marenzoni} (\bibinfo{year}{1995}):
  \emph{\bibinfo{title}{Parallel state space exploration for {GSPN} models}}.
\newblock In \bibinfo{editor}{G.~de~Michelis} \& \bibinfo{editor}{M.~Diaz},
  editors: {\sl \bibinfo{booktitle}{Applications and Theory of Petri Nets
  1995}}, {\sl \bibinfo{series}{LNCS}} \bibinfo{volume}{935},
  \bibinfo{publisher}{Springer Verlag}, pp. \bibinfo{pages}{181--200},
  \doi{10.1007/3-540-60029-9\_40}.

\bibitemdeclare{inproceedings}{cerna03distributed}
\bibitem{cerna03distributed}
\bibinfo{author}{I.~{\v{C}}ern\'{a}} \& \bibinfo{author}{R.~Pel\'{a}nek}
  (\bibinfo{year}{2003}): \emph{\bibinfo{title}{Distributed Explicit Fair Cycle
  Detection}}.
\newblock In: {\sl \bibinfo{booktitle}{Model Checking Software, 10th
  International SPIN Workshop}}, {\sl \bibinfo{series}{LNCS}}
  \bibinfo{volume}{2648}, \bibinfo{publisher}{Springer}, pp.
  \bibinfo{pages}{49--73}, \doi{10.1007/3-540-44829-2\_4}.

\bibitemdeclare{inproceedings}{cerna02relating}
\bibitem{cerna02relating}
\bibinfo{author}{I.~{\v{C}}ern\'a} \& \bibinfo{author}{R.~Pel\'anek}
  (\bibinfo{year}{2003}): \emph{\bibinfo{title}{Relating Hierarchy of Temporal
  Properties to Model Checking}}.
\newblock In: {\sl \bibinfo{booktitle}{Mathematical Foundations of Computer
  Science (MFCS)}}, {\sl \bibinfo{series}{LNCS}} \bibinfo{volume}{2747},
  \bibinfo{publisher}{Springer}, pp. \bibinfo{pages}{318--327},
  \doi{10.1007/978-3-540-45138-9\_26}.

\bibitemdeclare{inproceedings}{Vitter95}
\bibitem{Vitter95}
\bibinfo{author}{Yi-Jen Chiang}, \bibinfo{author}{Michael~T. Goodrich},
  \bibinfo{author}{Edward~F. Grove}, \bibinfo{author}{Roberto Tamassia},
  \bibinfo{author}{Darren~Erik Vengroff} \& \bibinfo{author}{Jeffrey~Scott
  Vitter} (\bibinfo{year}{1995}): \emph{\bibinfo{title}{External-memory graph
  algorithms}}.
\newblock In: {\sl \bibinfo{booktitle}{soda}}, \bibinfo{publisher}{Society for
  Industrial and Applied Mathematics}, pp. \bibinfo{pages}{139--149},
  \doi{10.1145/313651.313681}.

\bibitemdeclare{article}{ciardo98distributed}
\bibitem{ciardo98distributed}
\bibinfo{author}{G.~Ciardo}, \bibinfo{author}{J.~Gluckman} \&
  \bibinfo{author}{D.M. Nicol} (\bibinfo{year}{1998}):
  \emph{\bibinfo{title}{Distributed {S}tate {S}pace {G}eneration of
  {D}iscrete-{S}tate {S}tochastic {M}odels}}.
\newblock {\sl \bibinfo{journal}{INFORMS Journal on Computing}}
  \bibinfo{volume}{10}(\bibinfo{number}{1}), pp. \bibinfo{pages}{82--93},
  \doi{10.1287/ijoc.10.1.82}.

\bibitemdeclare{article}{ciardo97automated}
\bibitem{ciardo97automated}
\bibinfo{author}{Gianfranco Ciardo} (\bibinfo{year}{1997}):
  \emph{\bibinfo{title}{Automated parallelization of discrete state-space
  generation}}.
\newblock {\sl \bibinfo{journal}{J. Parallel Distrib. Comput.}}
  \bibinfo{volume}{47}, pp. \bibinfo{pages}{153--167},
  \doi{10.1006/jpdc.1997.1409}.

\bibitemdeclare{article}{Symmetry1}
\bibitem{Symmetry1}
\bibinfo{author}{E.~M. Clarke}, \bibinfo{author}{R.~Enders},
  \bibinfo{author}{T.~Filkorn} \& \bibinfo{author}{S.~Jha}
  (\bibinfo{year}{1996}): \emph{\bibinfo{title}{Exploiting symmetry in temporal
  logic model checking}}.
\newblock {\sl \bibinfo{journal}{Form. Methods Syst. Des.}}
  \bibinfo{volume}{9}(\bibinfo{number}{1-2}), pp. \bibinfo{pages}{77--104},
  \doi{10.1007/BF00625969}.

\bibitemdeclare{inproceedings}{clarke01progress}
\bibitem{clarke01progress}
\bibinfo{author}{E.M. Clarke}, \bibinfo{author}{O.~Grumberg},
  \bibinfo{author}{S.~Jha}, \bibinfo{author}{Y.~Lu} \&
  \bibinfo{author}{H.~Veith} (\bibinfo{year}{2001}):
  \emph{\bibinfo{title}{Progress on the {S}tate {E}xplosion {P}roblem in
  {M}odel {C}hecking}}.
\newblock In \bibinfo{editor}{R.~Wilhelm}, editor: {\sl
  \bibinfo{booktitle}{Informatics - 10 Years Back. 10 Years Ahead}}, {\sl
  \bibinfo{series}{LNCS}} \bibinfo{volume}{2000},
  \bibinfo{publisher}{Springer}, pp. \bibinfo{pages}{176--194}.

\bibitemdeclare{misc}{CUDA}
\bibitem{CUDA}
 (\bibinfo{year}{2009}): \emph{\bibinfo{title}{{NVIDIA CUDA Compute Unified
  Device Architecture - Programming Guide Version 2.0,}}}.
\newblock \bibinfo{note}{\url{http://www.nvidia.com/object/cuda\_develop.html},
  June 2009}.

\bibitemdeclare{inproceedings}{dill96themurphi}
\bibitem{dill96themurphi}
\bibinfo{author}{David~L. Dill} (\bibinfo{year}{1996}):
  \emph{\bibinfo{title}{The Mur$\varphi$ verification system}}.
\newblock In: {\sl \bibinfo{booktitle}{Conference on Computer-Aided
  Verification (CAV'96)}}, \bibinfo{series}{LNCS},
  \bibinfo{publisher}{Springer-Verlag}, pp. \bibinfo{pages}{390--393}.
\newblock \urlprefix\url{http://dl.acm.org/citation.cfm?id=647765.735832}.

\bibitemdeclare{inproceedings}{Semiext}
\bibitem{Semiext}
\bibinfo{author}{Stefan Edelkamp}, \bibinfo{author}{Peter Sanders} \&
  \bibinfo{author}{Pavel \v{S}ime\v{c}ek} (\bibinfo{year}{2008}):
  \emph{\bibinfo{title}{{Semi-external LTL Model Checking}}}.
\newblock In: {\sl \bibinfo{booktitle}{Computer Aided Verification (CAV
  2008)}}, \bibinfo{publisher}{Springer}, \bibinfo{address}{Berlin,
  Heidelberg}, pp. \bibinfo{pages}{530--542},
  \doi{10.1007/978-3-540-70545-1\_50}.

\bibitemdeclare{techreport}{ES08}
\bibitem{ES08}
\bibinfo{author}{Stefan Edelkamp} \& \bibinfo{author}{Damian Sulewski}
  (\bibinfo{year}{2008}): \emph{\bibinfo{title}{{Model Checking via Delayed
  Duplicate Detection on the GPU.}}}
\newblock \bibinfo{type}{Technical Report} \bibinfo{number}{Technical Report
  821}, \bibinfo{institution}{TU Dortmund}.
\newblock \urlprefix\url{http://www.tzi.de/~edelkamp/GPU\_Technical.pdf}.
\newblock \bibinfo{note}{Presented on the 22nd Workshop on Planning,
  Scheduling, and Design PUK 2008}.

\bibitemdeclare{misc}{SOCS09}
\bibitem{SOCS09}
\bibinfo{author}{Stefan Edelkamp} \& \bibinfo{author}{Damian Sulewski}
  (\bibinfo{year}{2009}): \emph{\bibinfo{title}{{Parallel State Space Search on
  the GPU}}}.
\newblock
  \urlprefix\url{http://www.cs.ualberta.ca/~nathanst/sara/papers/socs09\_submi%
ssion\_24.pdf}.
\newblock \bibinfo{note}{International Symposium on Combinatorial Search (SoCS
  2009)}.

\bibitemdeclare{inproceedings}{eicken92active}
\bibitem{eicken92active}
\bibinfo{author}{T.~von Eicken}, \bibinfo{author}{D.~E. Culler},
  \bibinfo{author}{S.~C. Goldstein} \& \bibinfo{author}{K.~E. Schauser}
  (\bibinfo{year}{1992}): \emph{\bibinfo{title}{Active messages: a mechanism
  for integrated communication and computation}}.
\newblock In: {\sl \bibinfo{booktitle}{19th Annual International Symposium on
  Computer Architecture}}, pp. \bibinfo{pages}{256--266},
  \doi{10.1145/285930.286002}.

\bibitemdeclare{article}{Symmetry2}
\bibitem{Symmetry2}
\bibinfo{author}{E.~Allen Emerson} \& \bibinfo{author}{A.~Prasad Sistla}
  (\bibinfo{year}{1996}): \emph{\bibinfo{title}{Symmetry and model checking}}.
\newblock {\sl \bibinfo{journal}{Form. Methods Syst. Des.}}
  \bibinfo{volume}{9}(\bibinfo{number}{1-2}), pp. \bibinfo{pages}{105--131},
  \doi{10.1007/BF00625970}.

\bibitemdeclare{inproceedings}{Evangelista08}
\bibitem{Evangelista08}
\bibinfo{author}{Sami Evangelista} (\bibinfo{year}{2008}):
  \emph{\bibinfo{title}{Dynamic Delayed Duplicate Detection for External Memory
  Model Checking}}.
\newblock In: {\sl \bibinfo{booktitle}{SPIN '08: Proc. of the 15th
  international workshop on Model Checking Software}},
  \bibinfo{publisher}{Springer}, \bibinfo{address}{Berlin, Heidelberg}, pp.
  \bibinfo{pages}{77--94}, \doi{10.1007/978-3-540-85114-1\_8}.

\bibitemdeclare{inproceedings}{garavel01parallel}
\bibitem{garavel01parallel}
\bibinfo{author}{H.~Garavel}, \bibinfo{author}{R.~Mateescu} \&
  \bibinfo{author}{I.M Smarandache} (\bibinfo{year}{2001}):
  \emph{\bibinfo{title}{Parallel {S}tate {S}pace {C}onstruction for
  {M}odel-{C}hecking}}.
\newblock In \bibinfo{editor}{Matthew~B. Dwyer}, editor: {\sl
  \bibinfo{booktitle}{Model Checking of Software (SPIN'2001)}}, {\sl
  \bibinfo{series}{LNCS}} \bibinfo{volume}{2057},
  \bibinfo{publisher}{Springer-Verlag}, pp. \bibinfo{pages}{216--234},
  \doi{10.1007/3-540-45139-0\_14}.

\bibitemdeclare{inproceedings}{geldenhuys99runtime}
\bibitem{geldenhuys99runtime}
\bibinfo{author}{Jaco Geldenhuys} \& \bibinfo{author}{P.~J.~A. de~Villiers}
  (\bibinfo{year}{1999}): \emph{\bibinfo{title}{Runtime Efficient State
  Compaction in {SPIN}}}.
\newblock In: {\sl \bibinfo{booktitle}{{SPIN}}}, pp. \bibinfo{pages}{12--21},
  \doi{10.1007/3-540-48234-2\_2}.

\bibitemdeclare{inproceedings}{govindaraju06gputerasort}
\bibitem{govindaraju06gputerasort}
\bibinfo{author}{Naga~K. Govindaraju}, \bibinfo{author}{Jim Gray},
  \bibinfo{author}{Ritesh Kumar} \& \bibinfo{author}{Dinesh Manocha}
  (\bibinfo{year}{2006}): \emph{\bibinfo{title}{{GPUTeraSort: high performance
  graphics co-processor sorting for large database management}}}.
\newblock In: {\sl \bibinfo{booktitle}{International Conference on Management
  of Data (SIGMOD 06)}}, \bibinfo{publisher}{ACM}, pp.
  \bibinfo{pages}{325--336}, \doi{10.1145/1142473.1142511}.

\bibitemdeclare{inproceedings}{Weber06}
\bibitem{Weber06}
\bibinfo{author}{Moritz Hammer} \& \bibinfo{author}{Michael Weber}
  (\bibinfo{year}{2006}): \emph{\bibinfo{title}{{"To Store or Not To Store"
  Reloaded: Reclaiming Memory on Demand}}}.
\newblock In: {\sl \bibinfo{booktitle}{Formal Methods: Applications and
  Technology}}, {\sl \bibinfo{series}{LNCS}} \bibinfo{volume}{4346},
  \bibinfo{publisher}{Springer}, pp. \bibinfo{pages}{51--66},
  \doi{10.1007/978-3-540-70952-7\_4}.

\bibitemdeclare{inproceedings}{HarishN07}
\bibitem{HarishN07}
\bibinfo{author}{P.~Harish} \& \bibinfo{author}{P.~J. Narayanan}
  (\bibinfo{year}{2007}): \emph{\bibinfo{title}{{Accelerating Large Graph
  Algorithms on the GPU Using CUDA}}}.
\newblock In: {\sl \bibinfo{booktitle}{HiPC}}, {\sl \bibinfo{series}{LNCS}}
  \bibinfo{volume}{4873}, \bibinfo{publisher}{Springer}, pp.
  \bibinfo{pages}{197--208}, \doi{10.1007/978-3-540-77220-0\_21}.

\bibitemdeclare{techreport}{harish2009large}
\bibitem{harish2009large}
\bibinfo{author}{P.~Harish}, \bibinfo{author}{V.~Vineet} \&
  \bibinfo{author}{P.~J. Narayanan} (\bibinfo{year}{2009}):
  \emph{\bibinfo{title}{{Large Graph Algorithms for Massively Multithreaded
  Architectures}}}.
\newblock \bibinfo{type}{Technical Report} \bibinfo{number}{IIIT/TR/2009/74},
  \bibinfo{institution}{Center for Visual Information Technology, International
  Institute of Information Technology Hyderabad, INDIA}.
\newblock
  \urlprefix\url{http://cvit.iiit.ac.in/papers/pawan09GraphAlgorithms.pdf}.

\bibitemdeclare{misc}{cuda_reduce}
\bibitem{cuda_reduce}
\bibinfo{author}{M.~Harris}: \emph{\bibinfo{title}{{Optimizing Parallel
  Reduction in CUDA,}}}.
\newblock
  \bibinfo{note}{\url{http://developer.download.nvidia.com/compute/cuda/1\_1/W%
ebsite/projects/reduction/doc/reduction.pdf}, March 2010}.

\bibitemdeclare{inproceedings}{haverkort99ontheefficient}
\bibitem{haverkort99ontheefficient}
\bibinfo{author}{B.~R. Haverkort}, \bibinfo{author}{A.~Bell} \&
  \bibinfo{author}{H.~C. Bohnenkamp} (\bibinfo{year}{1999}):
  \emph{\bibinfo{title}{{On the efficient sequential and distributed generation
  of very large Markov chains from stochastic Petri nets.}}}
\newblock In: {\sl \bibinfo{booktitle}{Petri Net and Performance Models
  (PNPM'99)}}, \bibinfo{publisher}{IEEE Computer Society Press}, pp.
  \bibinfo{pages}{12--21}, \doi{10.1109/PNPM.1999.796528}.

\bibitemdeclare{inproceedings}{haverkort98efficiency}
\bibitem{haverkort98efficiency}
\bibinfo{author}{Boudewijn~R. Haverkort}, \bibinfo{author}{Henrik Bohnenkamp}
  \& \bibinfo{author}{Alexander Bell} (\bibinfo{year}{1998}):
  \emph{\bibinfo{title}{{Efficiency Improvements in the Evaluation of Large
  Stochastic Petri Nets}}}.
\newblock In: {\sl \bibinfo{booktitle}{Forschungsbericht: 5. Workshop
  Algorithmen und Werkzeuge für Petrinetze}}, \bibinfo{publisher}{Universität
  Dortmund, Fachbereich Informatik}, pp. \bibinfo{pages}{55--61}.
\newblock
  \urlprefix\url{http://citeseerx.ist.psu.edu/viewdoc/download?doi=10.1.1.41.3%
435&rep=rep1&type=pdf}.

\bibitemdeclare{inproceedings}{Heljanko02parallelisation}
\bibitem{Heljanko02parallelisation}
\bibinfo{author}{Keijo Heljanko}, \bibinfo{author}{Victor Khomenko} \&
  \bibinfo{author}{Maciej Koutny} (\bibinfo{year}{2002}):
  \emph{\bibinfo{title}{{Parallelisation of the Petri Net Unfolding
  Algorithm}}}.
\newblock In: {\sl \bibinfo{booktitle}{Tools and Algorithms for the
  Construction and Analysis of Systems (TACAS 2002)}}, {\sl
  \bibinfo{series}{LNCS}} \bibinfo{volume}{2280},
  \bibinfo{publisher}{Springer}, pp. \bibinfo{pages}{371--385},
  \doi{10.1007/3-540-46002-0\_26}.

\bibitemdeclare{book}{spinbook}
\bibitem{spinbook}
\bibinfo{author}{Gerard~J. Holzmann} (\bibinfo{year}{2003}):
  \emph{\bibinfo{title}{{The Spin Model Checker: Primer and Reference
  Manual}}}.
\newblock \bibinfo{publisher}{Addison-Wesley}.

\bibitemdeclare{article}{Hol08}
\bibitem{Hol08}
\bibinfo{author}{Gerard~J. Holzmann} (\bibinfo{year}{2008}):
  \emph{\bibinfo{title}{{A Stack-Slicing Algorithm for Multi-Core Model
  Checking}}}.
\newblock {\sl \bibinfo{journal}{ENTCS}}
  \bibinfo{volume}{198}(\bibinfo{number}{1}), pp. \bibinfo{pages}{3--16},
  \doi{10.1016/j.entcs.2007.10.017}.

\bibitemdeclare{article}{holzmann07design}
\bibitem{holzmann07design}
\bibinfo{author}{Gerard~J. Holzmann} \& \bibinfo{author}{Dragan Bosnacki}
  (\bibinfo{year}{2007}): \emph{\bibinfo{title}{The Design of a Multicore
  Extension of the SPIN Model Checker}}.
\newblock {\sl \bibinfo{journal}{IEEE Trans. Software Eng.}}
  \bibinfo{volume}{33}(\bibinfo{number}{10}), pp. \bibinfo{pages}{659--674},
  \doi{10.1109/TSE.2007.70724}.

\bibitemdeclare{misc}{ifest}
\bibitem{ifest}
\emph{\bibinfo{title}{iFEST: Integration Framework for Embedded Systems
  Tools}}.
\newblock \urlprefix\url{http://www.artemis-ifest.eu}.

\bibitemdeclare{article}{InggsB05}
\bibitem{InggsB05}
\bibinfo{author}{Cornelia~P. Inggs} \& \bibinfo{author}{Howard Barringer}
  (\bibinfo{year}{2005}): \emph{\bibinfo{title}{{CTL$^*$ Model Checking on a
  Shared-Memory Architecture}}}.
\newblock {\sl \bibinfo{journal}{Electronic Notes in Theoretical Computer
  Science}} \bibinfo{volume}{128}(\bibinfo{number}{3}), pp.
  \bibinfo{pages}{107--123}, \doi{10.1016/j.entcs.2004.10.022}.

\bibitemdeclare{inproceedings}{JabbarE05}
\bibitem{JabbarE05}
\bibinfo{author}{Shahid Jabbar} \& \bibinfo{author}{Stefan Edelkamp}
  (\bibinfo{year}{2005}): \emph{\bibinfo{title}{{I/O Efficient Directed Model
  Checking}}}.
\newblock In: {\sl \bibinfo{booktitle}{Verification, Model Checking, and
  Abstract Interpretation (VMCAI 2005)}}, {\sl \bibinfo{series}{LNCS}}
  \bibinfo{volume}{3385}, \bibinfo{publisher}{Springer}, pp.
  \bibinfo{pages}{313--329}, \doi{10.1007/978-3-540-30579-8\_21}.

\bibitemdeclare{inproceedings}{JabbarE06}
\bibitem{JabbarE06}
\bibinfo{author}{Shahid Jabbar} \& \bibinfo{author}{Stefan Edelkamp}
  (\bibinfo{year}{2006}): \emph{\bibinfo{title}{{Parallel External Directed
  Model Checking with Linear I/O}}}.
\newblock In: {\sl \bibinfo{booktitle}{Verification, Model Checking, and
  Abstract Interpretation (VMCAI 2006)}}, {\sl \bibinfo{series}{LNCS}}
  \bibinfo{volume}{3855}, \bibinfo{publisher}{Springer}, pp.
  \bibinfo{pages}{237--251}, \doi{10.1007/11609773\_16}.

\bibitemdeclare{article}{Jayachandran06using}
\bibitem{Jayachandran06using}
\bibinfo{author}{G.~Jayachandran}, \bibinfo{author}{V.~Vishal} \&
  \bibinfo{author}{V.~S. Pande} (\bibinfo{year}{2006}):
  \emph{\bibinfo{title}{{Using massively parallel simulations and Markovian
  models to study protein folding: examining the villin head-piece}}}.
\newblock {\sl \bibinfo{journal}{Journal of Chemical Physics}}
  \bibinfo{volume}{124}(\bibinfo{number}{6}), p. \bibinfo{pages}{903–914},
  \doi{10.1063/1.2186317}.

\bibitemdeclare{inproceedings}{JonesM04}
\bibitem{JonesM04}
\bibinfo{author}{Michael Jones} \& \bibinfo{author}{Eric Mercer}
  (\bibinfo{year}{2004}): \emph{\bibinfo{title}{{Explicit State Model Checking
  with Hopper}}}.
\newblock In: {\sl \bibinfo{booktitle}{Model Checking Software (SPIN 2004)}},
  {\sl \bibinfo{series}{Lecture Notes in Computer Science}}
  \bibinfo{volume}{2989}, \bibinfo{publisher}{Springer}, pp.
  \bibinfo{pages}{146--150}, \doi{10.1007/978-3-540-24732-6\_10}.

\bibitemdeclare{article}{kahn62}
\bibitem{kahn62}
\bibinfo{author}{A.~B. Kahn} (\bibinfo{year}{1962}):
  \emph{\bibinfo{title}{Topological sorting of large networks}}.
\newblock {\sl \bibinfo{journal}{Communications of the ACM}}
  \bibinfo{volume}{5}(\bibinfo{number}{11}), pp. \bibinfo{pages}{558--562}.

\bibitemdeclare{article}{knottenbelt00probabilistic}
\bibitem{knottenbelt00probabilistic}
\bibinfo{author}{W.~Knottenbelt}, \bibinfo{author}{P.G Harrison},
  \bibinfo{author}{M.~Mestern} \& \bibinfo{author}{P.S. Kritzinger}
  (\bibinfo{year}{2000}): \emph{\bibinfo{title}{{A Probabilistic Dynamic
  Technique for the Distributed Generation of Very Large State Spaces}}}.
\newblock {\sl \bibinfo{journal}{Performance Evaluation}}
  \bibinfo{volume}{35}(\bibinfo{number}{1--4}), pp. \bibinfo{pages}{127--148},
  \doi{10.1016/S0166-5316(99)00061-9}.

\bibitemdeclare{inproceedings}{knottenbelt98probability}
\bibitem{knottenbelt98probability}
\bibinfo{author}{W.~Knottenbelt}, \bibinfo{author}{M.~Mestern},
  \bibinfo{author}{P.G Harrison},  \& \bibinfo{author}{P.S. Kritzinger}
  (\bibinfo{year}{1998}): \emph{\bibinfo{title}{Probability, Parallelism and
  the State Space Exploration Problem}}.
\newblock In \bibinfo{editor}{R.~Puigjaner}, editor: {\sl
  \bibinfo{booktitle}{Tools'98}}, {\sl \bibinfo{series}{LNCS}}
  \bibinfo{volume}{1469}, \bibinfo{publisher}{Springer Verlag}, pp.
  \bibinfo{pages}{165--179}, \doi{10.1007/3-540-68061-6\_14}.

\bibitemdeclare{inproceedings}{Korf04}
\bibitem{Korf04}
\bibinfo{author}{R.~Korf} (\bibinfo{year}{2004}):
  \emph{\bibinfo{title}{{Best-First Frontier Search with Delayed Duplicate
  Detection}}}.
\newblock In: {\sl \bibinfo{booktitle}{AAAI'04}}, \bibinfo{publisher}{AAAI
  Press / The MIT Press}, pp. \bibinfo{pages}{650--657}.
\newblock \urlprefix\url{http://dl.acm.org/citation.cfm?id=1597148.1597253}.

\bibitemdeclare{inproceedings}{KorfS05}
\bibitem{KorfS05}
\bibinfo{author}{R.~Korf} \& \bibinfo{author}{P.~Schultze}
  (\bibinfo{year}{2005}): \emph{\bibinfo{title}{{Large-Scale Parallel
  Breadth-First Search}}}.
\newblock In: {\sl \bibinfo{booktitle}{Proceedings of the 20th national
  conference on Artificial intelligence - Volume 3}}, \bibinfo{publisher}{AAAI
  Press / The MIT Press}, pp. \bibinfo{pages}{1380--1385}.
\newblock \urlprefix\url{http://dl.acm.org/citation.cfm?id=1619499.1619555}.

\bibitemdeclare{article}{KumarM05}
\bibitem{KumarM05}
\bibinfo{author}{Rahul Kumar} \& \bibinfo{author}{Eric~G. Mercer}
  (\bibinfo{year}{2005}): \emph{\bibinfo{title}{{Load Balancing Parallel
  Explicit State Model Checking}}}.
\newblock {\sl \bibinfo{journal}{Electronic Notes in Theoretical Computer
  Science}} \bibinfo{volume}{128}(\bibinfo{number}{3}), pp.
  \bibinfo{pages}{19--34}, \doi{10.1016/j.entcs.2004.10.016}.

\bibitemdeclare{inproceedings}{Laarman10boosting}
\bibitem{Laarman10boosting}
\bibinfo{author}{Alfons Laarman}, \bibinfo{author}{Jaco van~de Pol} \&
  \bibinfo{author}{Michael Weber} (\bibinfo{year}{2010}):
  \emph{\bibinfo{title}{{Boosting Multi-Core Reachability Performance with
  Shared Hash Tables}}}.
\newblock In: {\sl \bibinfo{booktitle}{Formal Methods in Computer-Aided Design
  (FMCAD 2010)}}, \bibinfo{publisher}{IEEE}, pp. \bibinfo{pages}{247--255}.
\newblock
  \urlprefix\url{http://ieeexplore.ieee.org/xpls/abs\_all.jsp?arnumber=5770956%
}.

\bibitemdeclare{inproceedings}{Lamborn08}
\bibitem{Lamborn08}
\bibinfo{author}{Peter Lamborn} \& \bibinfo{author}{Eric~A. Hansen}
  (\bibinfo{year}{2008}): \emph{\bibinfo{title}{Layered Duplicate Detection in
  External-Memory Model Checking}}.
\newblock In: {\sl \bibinfo{booktitle}{SPIN '08: Proc. of the 15th
  international workshop on Model Checking Software}},
  \bibinfo{publisher}{Springer}, \bibinfo{address}{Berlin, Heidelberg}, pp.
  \bibinfo{pages}{160--175}, \doi{10.1007/978-3-540-85114-1\_13}.

\bibitemdeclare{inproceedings}{lerda99distributed}
\bibitem{lerda99distributed}
\bibinfo{author}{Flavio Lerda} \& \bibinfo{author}{Riccardo Sisto}
  (\bibinfo{year}{1999}): \emph{\bibinfo{title}{Distributed-memory {M}odel
  {C}hecking with {SPIN}}}.
\newblock In: {\sl \bibinfo{booktitle}{Proc. of the 5th {I}nternational {SPIN}
  Workshop}}, {\sl \bibinfo{series}{LNCS}} \bibinfo{volume}{1680},
  \bibinfo{publisher}{Springer-Verlag}, pp. \bibinfo{pages}{22--39},
  \doi{10.1007/3-540-48234-2\_3}.

\bibitemdeclare{inproceedings}{lerda01addressing}
\bibitem{lerda01addressing}
\bibinfo{author}{Flavio Lerda} \& \bibinfo{author}{Willem Visser}
  (\bibinfo{year}{2001}): \emph{\bibinfo{title}{{Addressing Dynamic Issues of
  Program Model Checking}}}.
\newblock In: {\sl \bibinfo{booktitle}{International SPIN Workshop on Model
  Checking of Software (SPIN'2001)}}, {\sl \bibinfo{series}{LNCS}}
  \bibinfo{volume}{2057}, \bibinfo{publisher}{Springer}, pp.
  \bibinfo{pages}{80--102}, \doi{10.1007/3-540-45139-0\_6}.

\bibitemdeclare{proceedings}{ExternalMemory}
\bibitem{ExternalMemory}
\bibinfo{editor}{Ulrich Meyer}, \bibinfo{editor}{Peter Sanders} \&
  \bibinfo{editor}{Jop Sibeyn}, editors (\bibinfo{year}{2003}):
  \emph{\bibinfo{title}{Algorithms for Memory Hierarchies}}.
  \bibinfo{publisher}{Springer}.

\bibitemdeclare{inproceedings}{peled98ten}
\bibitem{peled98ten}
\bibinfo{author}{Doron Peled} (\bibinfo{year}{1998}): \emph{\bibinfo{title}{Ten
  Years of Partial Order Reduction}}.
\newblock In: {\sl \bibinfo{booktitle}{Proceedings of the 10th International
  Conference on Computer Aided Verification}},
  \bibinfo{publisher}{Springer-Verlag}, pp. \bibinfo{pages}{17--28}.

\bibitemdeclare{inproceedings}{pelanek09fighting}
\bibitem{pelanek09fighting}
\bibinfo{author}{R.~{Pelánek}} (\bibinfo{year}{2009}):
  \emph{\bibinfo{title}{{Fighting State Space Explosion: Review and
  Evaluation}}}.
\newblock In: {\sl \bibinfo{booktitle}{Formal Methods for Industrial Critical
  Systems (FMICS 2008)}}, {\sl \bibinfo{series}{LNCS}} \bibinfo{volume}{5596},
  \bibinfo{publisher}{Springer}, pp. \bibinfo{pages}{37--52},
  \doi{10.1007/978-3-642-03240-0\_7}.

\bibitemdeclare{inproceedings}{PennaITZ02}
\bibitem{PennaITZ02}
\bibinfo{author}{Giuseppe~Della Penna}, \bibinfo{author}{Benedetto Intrigila},
  \bibinfo{author}{Enrico Tronci} \& \bibinfo{author}{Marisa~Venturini Zilli}
  (\bibinfo{year}{2002}): \emph{\bibinfo{title}{{Exploiting Transition Locality
  in the Disk Based Mur$\varphi$ Verifier}}}.
\newblock In: {\sl \bibinfo{booktitle}{Formal Methods in Computer-Aided Design
  (FMCAD 2002)}}, pp. \bibinfo{pages}{202--219},
  \doi{10.1007/3-540-36126-X\_13}.

\bibitemdeclare{inproceedings}{stern97parallelizing}
\bibitem{stern97parallelizing}
\bibinfo{author}{U.~Stern} \& \bibinfo{author}{D.~L. Dill}
  (\bibinfo{year}{1997}): \emph{\bibinfo{title}{Parallelizing the Mur$\varphi$
  Verifier}}.
\newblock In \bibinfo{editor}{O.~Grumberg}, editor: {\sl
  \bibinfo{booktitle}{Proceedings of Computer Aided Verification ({CAV} '97)}},
  {\sl \bibinfo{series}{LNCS}} \bibinfo{volume}{1254},
  \bibinfo{publisher}{Springer-Verlag}, pp. \bibinfo{pages}{256--267},
  \doi{10.1007/BFb0028727}.

\bibitemdeclare{inproceedings}{Stern98}
\bibitem{Stern98}
\bibinfo{author}{U.~Stern} \& \bibinfo{author}{D.~L. Dill}
  (\bibinfo{year}{1998}): \emph{\bibinfo{title}{{Using Magnetic Disk Instead of
  Main Memory in the Mur$\varphi$ Verifier}}}.
\newblock In: {\sl \bibinfo{booktitle}{Computer Aided Verification. 10th
  International Conference}}, pp. \bibinfo{pages}{172--183},
  \doi{10.1007/BFb0028743}.

\bibitemdeclare{inproceedings}{vardi86autom}
\bibitem{vardi86autom}
\bibinfo{author}{M.Y. Vardi} \& \bibinfo{author}{P.~Wolper}
  (\bibinfo{year}{1986}): \emph{\bibinfo{title}{An automata-theoretic approach
  to automatic program verification}}.
\newblock In: {\sl \bibinfo{booktitle}{Proc. IEEE Symposium on Logic in
  Computer Science}}, \bibinfo{publisher}{Computer Society Press}, pp.
  \bibinfo{pages}{322--331}.

\bibitemdeclare{inproceedings}{VBBB09}
\bibitem{VBBB09}
\bibinfo{author}{K.~Verstoep}, \bibinfo{author}{H.~Bal},
  \bibinfo{author}{J.~Barnat} \& \bibinfo{author}{L.~Brim}
  (\bibinfo{year}{2009}): \emph{\bibinfo{title}{{Efficient Large-Scale Model
  Checking}}}.
\newblock In: {\sl \bibinfo{booktitle}{23rd IEEE International Parallel \&
  Distributed Processing Symposium (IPDPS 2009)}}, \bibinfo{publisher}{IEEE},
  pp. \bibinfo{pages}{1--12}, \doi{10.1109/IPDPS.2009.5161000}.

\bibitemdeclare{inproceedings}{ZhouH04a}
\bibitem{ZhouH04a}
\bibinfo{author}{Rong Zhou} \& \bibinfo{author}{Eric~A. Hansen}
  (\bibinfo{year}{2004}): \emph{\bibinfo{title}{{Structured Duplicate Detection
  in External-Memory Graph Search}}}.
\newblock In: {\sl \bibinfo{booktitle}{AAAI}}, \bibinfo{publisher}{AAAI Press /
  The MIT Press}, pp. \bibinfo{pages}{683--689}.
\newblock \urlprefix\url{http://www.aaai.org/Papers/AAAI/2004/AAAI04-108.pdf}.

\end{thebibliography}

\end{document}